\documentclass{article}
\usepackage{spconf,amsmath,graphicx}

\usepackage{cite}
\usepackage{amsmath,amssymb,amsfonts}
\usepackage{algorithmic}
\usepackage{graphicx}
\usepackage{textcomp}
\usepackage{xcolor}
\usepackage{comment}
\usepackage{soul}   
\usepackage{algorithm}
\usepackage{algorithmic} 
\usepackage{makecell} 
\usepackage{subfig} 
\graphicspath{{figures/}}
\usepackage{cleveref} 

\title{Memory Efficient Corner Detection for Event-driven Dynamic Vision Sensors}
\name{Pao-Sheng Vincent Sun*, Arren Glover\textsuperscript{+}, Chiara Bartolozzi\textsuperscript{+} and Arindam Basu*}
\address{City University of Hong Kong*\\
        Italian Institute of Technology\textsuperscript{+}\\
}
\begin{document}
%
\maketitle
\begin{abstract}

Event cameras offer low-latency and data compression for visual applications, through event-driven operation, that can be exploited for edge processing in tiny autonomous agents.  
Robust, accurate and low latency extraction of highly informative features such as corners is key for most visual processing. While several corner detection algorithms have been proposed, state-of-the-art performance is achieved by ``luvHarris''. However, this algorithm requires a high number of memory accesses per event, making it less-than ideal for low-latency, low-energy implementation in tiny edge processors. In this paper, we propose a new event-driven corner detection implementation tailored for edge computing devices, which requires much lower memory access than luvHarris while also improving accuracy. Our method trades computation for memory access, which is more expensive for large memories. For a DAVIS346 camera, our method requires $\approx 3.8 X$ less memory, $\approx 36.6 X$ less memory accesses with only $\approx 2.3X$ more computes. 
\end{abstract}
\section{Introduction}
\label{sec:intro}

Machine vision is an integral part of technological innovation, supporting autonomous operation of vehicles, Unmanned Aerial Vehicles (UAV), drones, robots, Internet of Things (IoT) and many other applications. Its advancements come from the improvement of both the hardware used and the algorithms that are run on the hardware. Frame-based cameras have been the default vision sensor till date. They capture the scene based on sampling light intensity at fixed exposure time. However, despite the abundance of information provided, the bandwidth required to transmit the data from the camera to the processor is enormous, since every pixel's information is transmitted regardless of their content. Their temporal resolution is limited by the exposure time, that causes also motion blur for fast moving stimuli. 
To overcome these bottlenecks, event cameras sample the visual signal only at fixed changes in temporal contrast\cite{tobi_dvs,posch_review}, removing redundancy due to unchanged input, so that their output bandwidth is significantly reduced as compared to frame-based cameras, featuring higher dynamic range, lower-latency, and lower power consumption. 

The spatio-temporal sparsity of events from stationary event cameras allows usage of Deep Neural Networks (DNNs) in edge IoT applications by enabling duty-cycling of the DNN processor\cite{xueyong_jssc,sumon_jssc}; however, this requires the availability of the results from low-level image processing steps such as noise removal and corner detection. Hence, these steps need to be computed with high energy efficiency since they reside in the always ON part of the system. Several algorithms and hardware implementations show promising results \cite{gallego2020event, vasco2016fast, mueggler2017fast, alzugaray2018asynchronous, glover2021luvharris}. Unfortunately, all of these were designed for using powerful processors consuming hundreds time more than what is available in extreme edge applications\cite{benini_edge1}. \cite{xueyong_jssc,sumon_jssc} demonstrate in-memory computing based noise removal with high energy efficiency; but corner detection methods are yet to be optimized for edge devices.

\begin{figure}[!ht]
    \centering
    \includegraphics[width=0.48\textwidth]{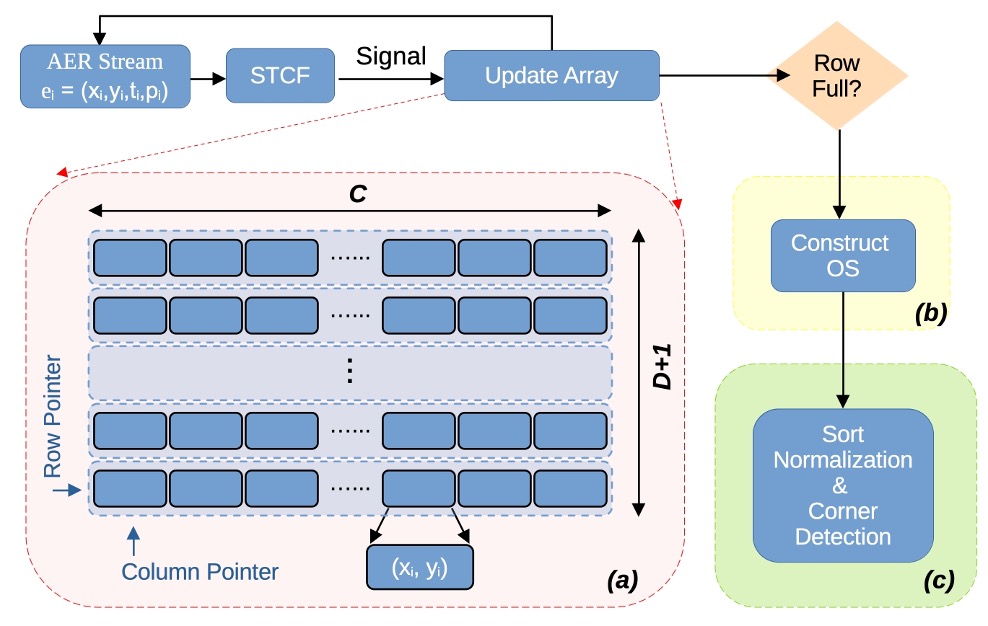}
    \caption{The algorithm presented in this paper can be broken down into three stages. a) Update 2D Array with temporal ordering b) Construct the OS which introduces relative spatial ordering. c) Perform corner detection after sort normalization on the current row.}
    \label{fig:overall_pipeline}
\end{figure}

Therefore, in this paper, we present a corner detection pipeline (Fig. \ref{fig:overall_pipeline}) that can be deployed on edge devices, featuring: 1) A memory efficient method for storing events while retaining the temporal ordering between events, 2) An efficient process to convert events into a fixed size matrix representation for corner detection; 3) A computationally efficient process for classifying events as corners without sacrificing accuracy. 


\section{Background}
\label{sec:background}

\subsection{Event Camera}


Event cameras feature sparse event outputs that encode change in illumination. These outputs are typically encoded using 
the $x$ and $y$ coordinate of the output pixel, the timestamp of the event, $t$, and the polarity of the event, $p$. The polarity of the event denotes a positive (ON event) or negative (OFF event) change in brightness. 


\subsection{Event Based Corner Detection}
Computer vision tasks need sharp, well-defined features transferable across different scenarios and tasks. One of the most popular and informative features to detect is a corner. Plenty of corner detectors have been developed, with techniques based on either change in luminous intensity, such as Harris corner detector \cite{harris1988combined}, or segment-based techniques like Features from accelerated segment test (FAST) \cite{rosten2006machine}.

These algorithms have also been adopted for event-based corner detection, with a data structure termed Surface of Active Events (SAE)\cite{benosman2013event} as a replacement for the image. For Harris-based corner detection, such as eHarris \cite{vasco2016fast, mueggler2017fast}, the Harris detector is applied per event to a patch extracted from a binarized SAE surrounding the event location. Its main bottleneck is the throughput, that scales with the number of events, limiting the applicability of this method in real-world scenarios. Segment-based algorithms like FAST \cite{mueggler2017fast} and ARC* \cite{alzugaray2018asynchronous} can be directly applied to a SAE, and perform calculations at a higher throughput, as no complex operations like matrix multiplication are needed, at the cost of limiting the accuracy of the detection. Despite this, their throughput for high resolution event cameras moving in cluttered scenes is still limited, and they are prone to false positives \cite{glover2021luvharris}. 

LuvHarris \cite{glover2021luvharris} attempts to increase the throughput while maintaining high accuracy by proposing two changes: 1) A variant of the SAE called Threshold-Ordinance Surface (TOS) which uses 8-bit unsigned integers to represent the newness of the events recorded, and 2) a pipeline that uses multi-threaded processing to improve upon the throughput and accuracy, with one thread used to update TOS every event while another thread updates the Harris score as a Lookup table (LUT) as fast as possible. \emph{While luvHarris produces state-of-the-art results in terms of accuracy, and increases throughput by having only a small fraction of the computation performed per event (i.e. the TOS update), it requires a per-event update of a neighbourhood of pixels in the TOS.} Unfortunately, the energy required for memory access dominates computation cost in modern VLSI processes \cite{horowitz}, and the TOS update becomes a memory bottleneck making the direct adoption of luvHarris difficult for ASICs designed for the extreme edge.


\section{Methods}
\label{sec:method}
We build on luvHarris, tailoring its implementation to hardware for edge computation. 

The first bottleneck of luvHarris comes in terms of updating the TOS for every event. Intuitively, the TOS in luvHarris maintains a global spatial memory of events which is ordered locally by the neighbourhood update. For each incoming event, a square patch of TOS around the pixel location of the event is updated, requiring xxx memory writes per event.  We propose next a data structure and update method to have a similar effect but without the numerous writes to memory. We use a 2D temporal array that is updated per event, and used to create an Ordered Surface, that is then used as input to the module computing the Harris score.

The computation of the Harris score requires two convolutions of an image patch with Sobel operators, element-wise multiplications to create $3$ intermediate matrices and finally smoothing of these $3$ matrices by a Gaussian kernel. Assuming typical values of a $7\times7$ patch and $3\times3$ Sobel operator, this results in $\approx 600$ MAC operations. With a $100$ MHz clock, this results in a throughput of only $166k$ corner evaluations per second (CEPS), which is far below event rates from modern event camera which easily reach $3-4$ MEPS\cite{glover2021luvharris}. luvHarris has a high event throughput by using a LUT to assign corner detections. The table is updated only as fast as hardware can allow. However, with the advent of In-memory computing (IMC) to accelerate convolutional neural networks (CNN)\cite{imc_cnn}, the Harris evaluation may be performed much more efficiently on ASICs. Using IMC, all convolutions (including Gaussian smoothing) can be done in parallel in $1$ cycle ($49\times25$ crossbar for Sobel and $25\times3$ crossbar for Gaussian, both of which are well within reported IMC macro sizes \cite{imc_cnn}) and need $25$ cycles for the $3$ intermediate matrices for a total of $27$ cycles. With the same clock speed, this results in a throughput of $\approx 3.7M$ CEPS, sufficient to match event camera event rates. \emph{Hence, we propose to revert to a low-memory-access, event-by-event Harris calculation for edge devices, while maintaining other advantages of the luvHarris algorithm.}

\subsection{2D Temporal Array}
To keep a continuous history of events that can be used to construct the Ordered Surface (OS) for corner detection, we propose a 2D array data structure (Fig. \ref{fig:overall_pipeline}) which is capable of maintaining global temporal ordering between events while distinctly marking which events are yet to be classified. The salient points about this structure are: 1) $D+1$ rows and $C$ columns; each cell stores the $(x,y)$ coordinates of the events; 2) A row pointer and column pointer to keep track of the insertion location. 3) $C$ determines how many events need to be accumulated before the construction of OS and corner detection stage is triggered--this can potentially increase accuracy over eHarris as fully event-by-event detection can have high-frequency noise; 4) $E_{hist}=D\times C$ determines the number of events in an OS used in corner detection.



We use a column pointer to keep track of the current insertion location and to determine if the construction of the OS and corner detection process can begin. An additional row ($D+1$-th) is used to allow insertion of events while Harris evaluation is performed on the OS. Events from the event camera will be sent through a Spatial-Temporal Correlation Filter (STCF) \cite{guo2022low} first to remove ``noise" events. If the event is classified as signal, it will be stored within the array. The column pointer is incremented on a per-event basis, with a circular reset mechanism, i.e. it will reset back to zero (the start of an array row) after it points to the end of the row and starts the construction of the OS and the corner detection process.

One of the flexibilities offered by this data structure is the ability to adjust the number of events available for corner detection. For increasing the number of rows, the event history $E_{hist}$ available increases, and more events are used for constructing the OS (longer history). However, an unconstrained increase in the event history size is unnecessary and costly, since events that are temporally further away from the current batch do not contribute significantly to the event, but could hinder the detection accuracy as unwanted noise. \emph{Evaluating in batches of $C$ events offers the advantage of providing more support events to the local patch when calculating the Harris response for the event of interest. Also, keeping this batch of $C$ events for $D$ more cycles allows for an overlapping stride over event history while 1D queues\cite{mueggler2017fast} have non-overlapping stride.}
At the same time, differently from luvHarris, this method relies on a fixed number of events for the computation, being less robust to sudden changes of clutter or accelerations that structurally modify the number of generated events. 
Nevertheless, for many robotic edge applications, such as navigation, appropriate parameter tuning can be found for a given environment.


\subsection{Ordered Surface}
Transforming events into an fixed size matrix representation introduces spatial information enabling conventional computer vision techniques to be applied directly. The proposed OS is designed to represent the latest $E_{hist}$ events as an image such that the global temporal order and spatial relationship of the events can be visualized. As the 2D array already maintains a temporal ordering of the different events, we can assign an order value to each event such that the more recent the event is, the larger its value will be. The 2D array is traversed from the row with the oldest events to the current row, as indicated by the row pointer. The process for generating the OS is summarised in \Cref{alg:OS}. The temporary OS in our algorithm replaces the TOS in luvHarris; but OS only requires $D$ write per event compared to  $(2k+1)^2$ write per event in luvHarris. \emph{While the OS has both spatial and temporal information, the temporal information is based on global ordering and hence spans a much larger range of value than the $8$ bits used in TOS.} This is corrected by sort normalization in the next step of patch creation. 

\begin{algorithm}[tb]
    \caption{Ordered Surface Construction}
    \label{alg:OS}
\begin{algorithmic}
    \REQUIRE $2D\ Array, rowPtr\ (row\ pointer), OS$
    \STATE $T,\ OS \gets 1,\ 0$
    \FOR{$r = (rowPtr+2)\bmod{(D+1)}:rowPtr$}
        \FOR{$c = 0:C$}
            \STATE $x,y \gets array_{r,c}$
            \STATE $OS_{x,y} \gets T$
            \STATE $T \gets T + 1$
        \ENDFOR
    \ENDFOR
\end{algorithmic}
\end{algorithm}

\subsection{Patch Creation and Corner Detection}
To take advantage of the sparsity of events, only a local patch surrounding the event of interest is extracted from the OS for the Harris calculation. The algorithm is presented in \Cref{alg:patch}. The global ordering information in the OS spans a large range of values ($0-C\times D$) which makes it difficult for Harris evaluation. Normalizing the global indices to span the range of patch indices would normally require costly division operation. \emph{Instead, the non-zero values in the patch are added to a queue and sorted in descending order.} Then based on the order, the values are normalized such that the normalized pixel value is equal to $255-index$, where $index$ refers to the pixel's position in the sorted queue--we refer to this process as sort normalization. From a theoretical point of view, our patch normalization is similar to rank order coding\cite{thorpe_rank} which has been postulated as one potential encoding strategy used in visual cortex. Once the patch is extracted and normalized, we can apply the Harris detector to it, generate the score for the event of interest, and classify it as a corner if the score exceeds a threshold. As mentioned earlier, we expect to use IMC approaches to accelerate the Harris evaluation. Also, we perform this calculation for all of the events stored within the latest row in a batch of $C$--this provides more context to correctly classify corner events.

\begin{algorithm}[!th]
    \caption{Local Patch Construction and Harris Calculation}
    \label{alg:patch}
\begin{algorithmic}
    \REQUIRE $x, y, OS, k\ (Patch\ Half\ Size)$
    \FOR{$r = -k:k$}
        \FOR{$c = -k:k$}
            \STATE $px,\ py \gets c+k,\ r+k$
            \STATE $Patch_{px,py} \gets OS_{x+c,y+r}$
        \ENDFOR
    \ENDFOR
    \STATE Sort $SortQueue$ in descending order
    \STATE $SortValue \gets 255$
    \FOR{$value : SortQueue$}
        \STATE $Patch_{Value} \gets SortValue$
        \STATE $SortValue \gets SortValue - 1$
    \ENDFOR
    \STATE $Score \gets HarrisDetector(Patch)$
    \IF{$Score >= Threshold$}
        \STATE $CornerFlag \gets True$
    \ELSE
        \STATE $CornerFlag \gets False$
    \ENDIF
\end{algorithmic}
\end{algorithm}

\section{Results}
\label{sec:results}

\subsection{Setup}
To compare the performance of the proposed algorithm against existing algorithm with the same detector, two different datasets from  \cite{mueggler2017event} are chosen. The method for generating the ground truth used to determine the accuracy of the algorithm is the one proposed in \cite{glover2021luvharris}, where the event file is converted into standard image with e2vid \cite{rebecq2019high, rebecq2019events} and the corresponding Harris score is generated from these images, with the assumption that the top 20\% scoring events are corner events.

The algorithms chosen for comparison are luvHarris \cite{glover2021luvharris} and eHarris \cite{vasco2016fast}. The Harris detector from the OpenCV library \cite{bradski2000opencv} is used for our algorithm (as in luvHarris). The sorting algorithm used comes from C++'s standard template library.

\subsection{Corner Detection Accuracy}
\begin{figure}[!t]
    \begin{minipage}[b]{0.5\linewidth}
        \centering
        \centerline{\includegraphics[width=1.0\linewidth]{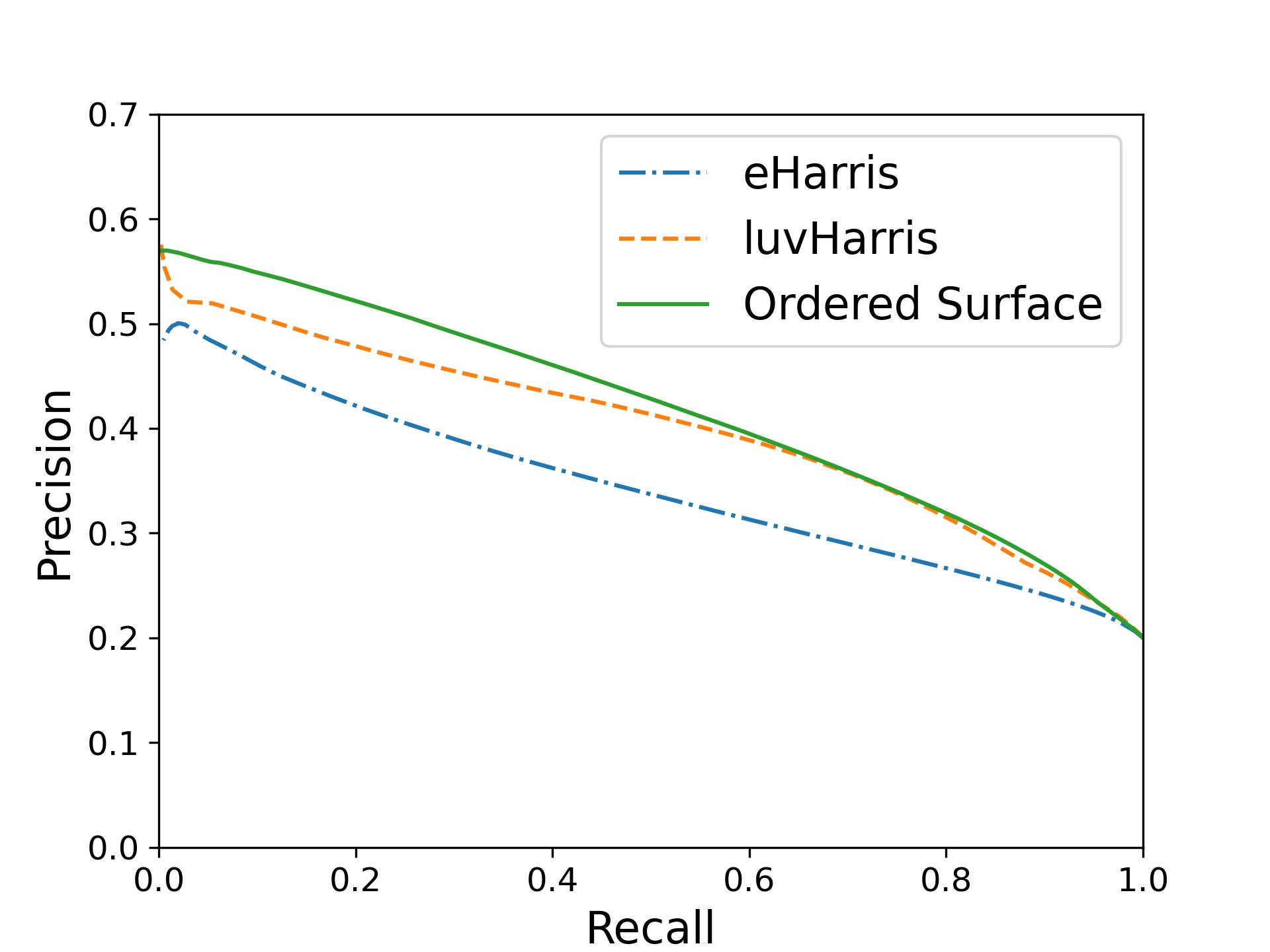}}
        \centerline{(a) Shapes 6DoF.}
    \end{minipage}%
    \begin{minipage}[b]{0.5\linewidth}
        \centering
        \centerline{\includegraphics[width=1.0\linewidth]{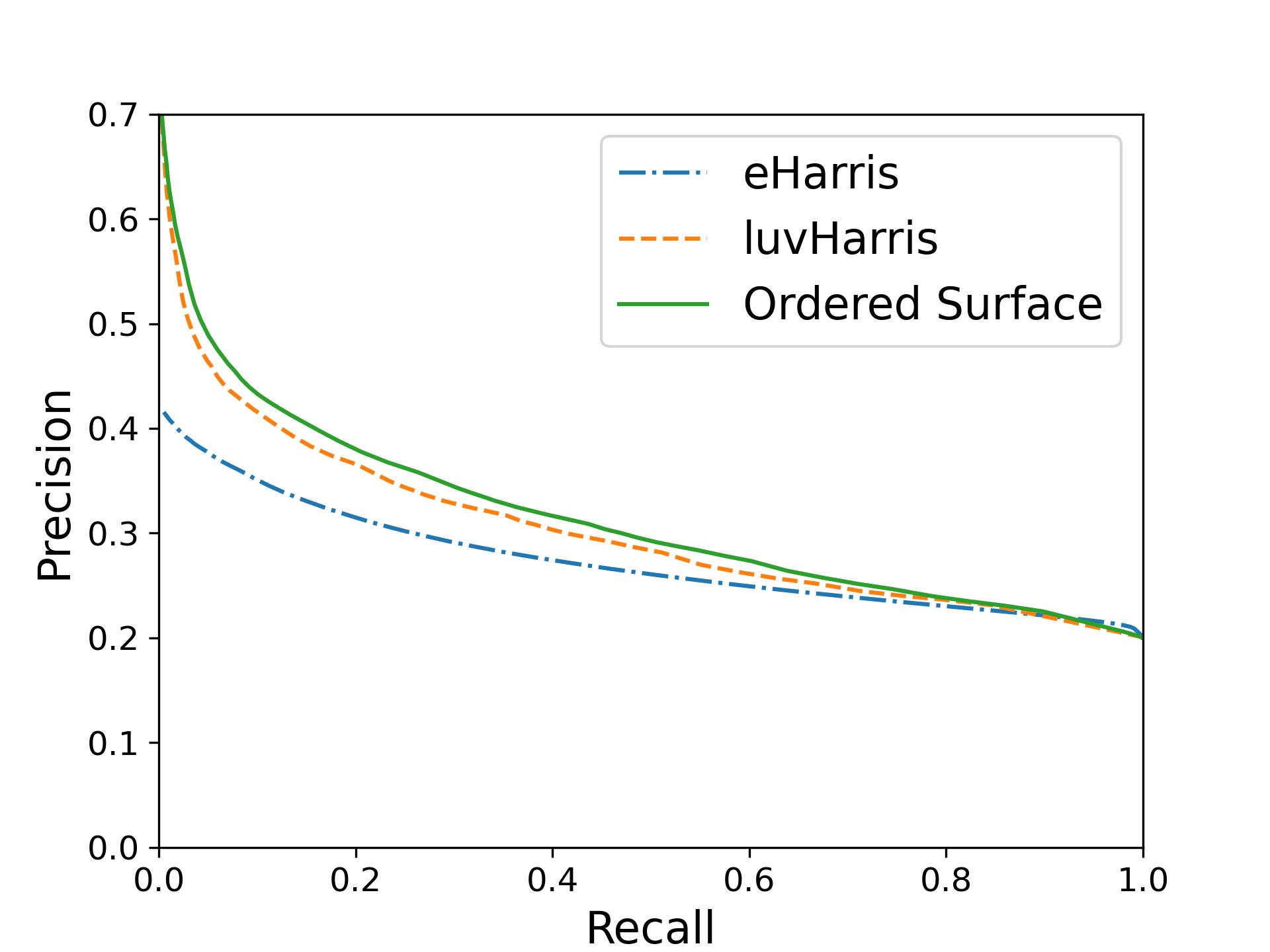}}
        \centerline{(b) Dynamic 6DoF.}
    \end{minipage}%
\caption{Precision Recall Plot for two files in the Event Camera Datasets \cite{mueggler2017event}.}
\label{fig:result_pr_plot}
\end{figure}

The performance of the proposed algorithm compared to luvHarris and eHarris can be seen in \Cref{fig:result_pr_plot}. For these simulations, $C=100$, $D=10$ or $16$, and $k=3$ where the patch for Harris calculation is of size $P=(2k+1)\times (2k+1)$. These parameters were selected based on a hyperparameter search across a small validation set. Similar to \cite{glover2021luvharris}, we sweep the threshold to generate ROC curves instead of comparing corner detection at a fixed threshold. As seen from the two subplots, our proposed algorithm outperforms eHarris by a large margin and is capable of matching or outperforming the accuracy obtained by luvHarris.


\subsection{Resource requirement}
We compare the resource requirement of our algorithm with the state-of-the-art (SOTA) luvHarris algorithm in terms on: 1) Number of operations per event; 2) Energy usage per event; 3) Memory required to run the algorithm. Intuitively, our method exchanges the extra memory access of updating neighbourhood pixels in TOS of luvHarris with sort operations ($(1-S)Plog_2((1-S)P)$ operations for sorting non-zero elements) during the patch normalization. Assuming the usage of a DAVIS346 event camera with resolution $346\times260$ and a conservative sparsity estimate of $S=0.5$ in the patches ($S=0.62$ and $0.8$ in the two datasets shown), our proposed method requires $\approx 2.3X$ more computations. However, the number of writes of an event to the OS memory in our method is only $D$ times during its lifetime in the 2D array. In comparison, $P=(2k+1)^2$ elements have to read and written back per event in the TOS memory. With a conservative estimate of write energy being $\approx 3X$ read energy of a SRAM (this can easily be 6-10X\cite{sumon_jssc_1}  for large SRAM), the memory access energy of our method is $\approx 36.6X$ less than luvHarris. Lastly, luvHarris maintains $8$-bit and $32$-bit per-pixel memories for the TOS and Harris LUT respectively while our method requires only an $10-11$-bit per pixel memory for the OS resulting in a $\geq 3.8X$ memory savings compared to luvHarris\cite{glover2021luvharris}.

\section{Conclusion and Discussion}
\label{sec:conclusion}

In this paper, we have proposed an algorithm for event based corner detection for edge devices, which uses less memory and memory accesses than SOTA algorithms. Despite the smaller memory footprint and energy usage, the algorithm still achieves accuracy comparable/better than SOTA algorithm like luvHarris. 
However, the Harris LUT in \cite{glover2021luvharris} provides an elegant method for embedded platforms with serial processors where the speed of Harris evaluation is a bottleneck. Our proposed work is intended for ASICs that use IMC to accelerate Harris evaluations in which, instead, memory access becomes the bottleneck. Future work will try to keep the Harris LUT for event-rate scalable implementations while using IMC for the TOS update since the TOS representation has the added advantage of not requiring tuning of event history according to application, with adaptability similar to \cite{li2023asynchronous}. Experiments to evaluate the actual throughput achieved is also required for a valid comparison of algorithms.

\bibliographystyle{IEEEtran}

\end{document}